\documentclass{article}
\usepackage{spconf,amsmath,graphicx,hyperref}

\usepackage[utf8]{inputenc}
\usepackage{longtable}
\usepackage{booktabs}
\usepackage{ragged2e}
\usepackage{tablefootnote}
\usepackage{array}
\usepackage{multirow}
\usepackage{graphicx}
\usepackage[polish,english]{babel}
\usepackage[T1]{fontenc}

\usepackage{url}

\title{Baseline Systems for the 2025 Low-Resource Audio Codec Challenge}
\name{Yusuf Ziya Isik, Rafał Łaganowski}
\address{Collaboration AI, Cisco Systems, Inc.}
\begin{document}
%
\maketitle
\begin{abstract}
The Low-Resource Audio Codec (LRAC) Challenge aims to advance neural audio coding for deployment in resource-constrained environments. The first edition focuses on low-resource neural speech codecs that must operate reliably under everyday noise and reverberation, while satisfying strict constraints on computational complexity, latency, and bitrate. Track 1 targets \textit{transparency codecs}, which aim to preserve the perceptual transparency of input speech under mild noise and reverberation. Track 2 addresses \textit{enhancement codecs}, which combine coding and compression with denoising and dereverberation. This paper presents the official baseline systems for both tracks in the 2025 LRAC Challenge. The baselines are convolutional neural codec models with Residual Vector Quantization, trained end-to-end using a combination of adversarial and reconstruction objectives. We detail the data filtering and augmentation strategies, model architectures, optimization procedures, and checkpoint selection criteria.
\end{abstract}
\section{Introduction}
\label{sec:intro}

This paper presents the design and training of the baseline models for the two tracks of the 2025 LRAC Challenge.\footnote{\url{https://lrac.short.gy/2025-lrac-challenge}} The challenge imposes strict constraints on latency, computational complexity, and transmission bandwidth. All participating codec systems must operate at a 24 kHz sampling rate and support both an ultra-low bitrate mode (up to 1 kbps) and a low-bitrate mode (up to 6 kbps) within a single system. Track 1, the transparency codec track, permits up to 30 ms of latency, including buffering but excluding processing latency. Track 2, the enhancement codec track, allows up to 50 ms of latency. The computational complexity limits are 700 MFLOPS for Track 1 (with no more than 300 MFLOPS on the receive side) and 2600 MFLOPS for Track 2 (with no more than 600 MFLOPS on the receive side). 

The baseline systems are designed to demonstrate codec implementations that meet the challenge constraints, provide a benchmark for participants, and facilitate entry into the competition. They are made available through two separate public repositories.

The LRAC data generation repository~\cite{lrac-data} contains scripts to download public datasets, apply preprocessing (such as sampling rate conversion), and curate data using pre-generated file lists. It also handles splitting the data into training, validation, and open test sets to use during the development phase. The actual test phase relies on a blind test set, which will be released at the end of the development phase.

The LRAC baseline development repository~\cite{lrac-baseline} is a public fork of the End-to-End Speech Processing (ESPnet) toolkit~\cite{watanabe2018espnet}. It enhances the existing GAN-based neural speech codec training recipes, providing greater flexibility in model and loss function design, and improves the vector quantization implementation. The repository includes designs and configurations for models, loss functions, data loaders, and optimizers. The trained baseline model weights are also provided in the repository under a Creative Commons Attribution-NonCommercial license.

It should be noted that these baseline neural codecs were developed exclusively for the 2025 LRAC Challenge and are not intended for, nor deployed in, any commercial products.

\section{DATASETS AND AUGMENTATIONS}
\label{sec:format}

To ensure fair comparison across submissions and to facilitate analysis of factors influencing system performance, the LRAC Challenge is conducted on a closed set of publicly available training data for both speech and noise. Publicly available room impulse responses (RIRs) are included in the training data; however, participants may additionally use other RIRs, either recorded or synthetically generated.

For the baseline systems, data preparation involves filtering a curated subset of the original speech files provided by Collaboration AI, Cisco Systems. File selection is guided by estimated quality metrics for signal-to-noise ratio (SNR), reverberation, and effective speech bandwidth. To promote diversity and balance, we further stratify the dataset according to speaker gender, speaker identity, and per-speaker recording durations, using ground-truth annotations when available or estimated values otherwise. Files reserved for the open test set are excluded from training, ensuring no speaker overlap between training and evaluation data. The baseline data generation recipe further sets aside part of the data as validation dataset to be used in hyperparameter tuning and checkpoint selection. Table~\ref{tab:lrac-training-data} summarizes the speech datasets and their total durations before and after curation.

\begin{table}[!t]
  \centering
  \footnotesize
      \caption{Training speech data curation by dataset.}
  \label{tab:lrac-training-data}
  \begin{tabular}{lrrr}
    \toprule
    \textbf{Dataset} & \textbf{Kept (h)} & \textbf{Original (h)} & \textbf{Retention} \\
    \midrule
    LibriTTS          & 46.3   & 191.3  & 24.20\% \\
    VCTK              & 22.3   & 78.8   & 28.30\% \\
    EARS              & 86.8  & 86.8  & 100.00\% \\
    Librivox (DNS5)   & 85.3   & 313.9  & 27.17\% \\
    MLS (FR, DE, ES)  & 275.6  & 450.0  & 61.24\% \\
    GLOBE             & 186.4  & 520.9  & 35.78\% \\
    \midrule
    \textbf{Total}    & \textbf{702.7} & \textbf{1641.7} & \textbf{42.80\%} \\
    \bottomrule
  \end{tabular}
\end{table}

\begin{figure}[!b]
\label{fig:lrac-noise-data}
\begin{minipage}[b]{1.0\linewidth}
  \centering
  \centerline{\includegraphics[width=8.5cm]{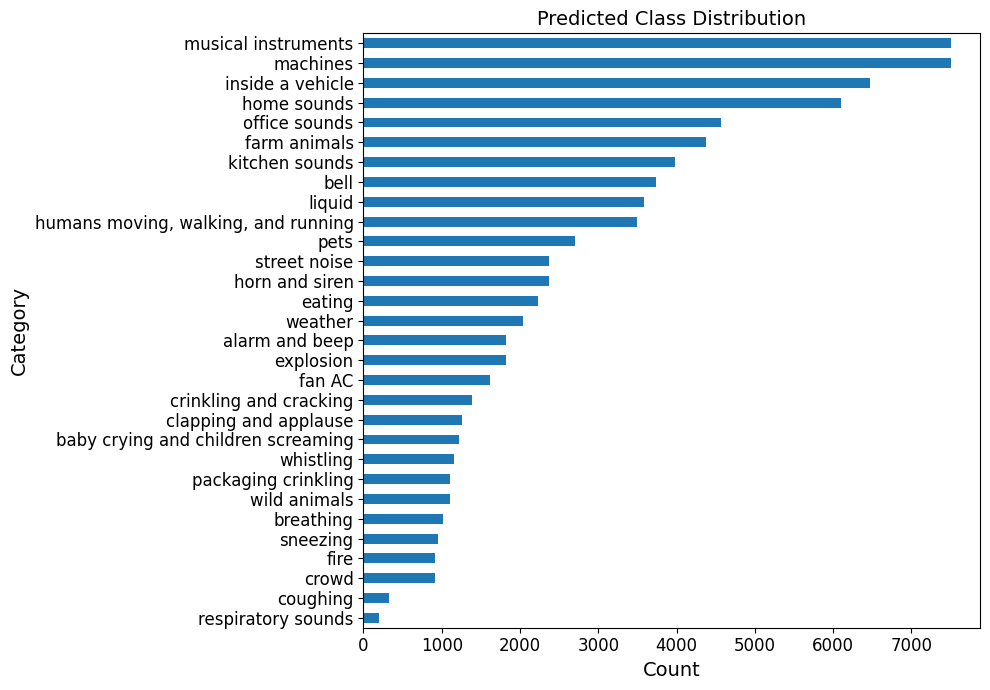}}
  \centerline{The final training noise data distribution.}\medskip
\end{minipage}
\end{figure}

The baseline data preparation pipeline also filters noise files based on the curated selection provided by Collaboration AI, Cisco Systems. This curation ensures a diverse and balanced noise dataset spanning major noise categories. To achieve this, we classify all noise files using an ontology derived from AudioSet, simplified to emphasize broad noise types and human vocal sounds. Noise classification is performed with CLAP~\cite{clap}, and files from the most frequent categories are downsampled to balance the distribution. A subset of the noise data is reserved for constructing the open test set used during development, while additional portions of the training noise and reverberation data are held out to form a validation set for hyperparameter tuning and checkpoint selection. Figure~\ref{fig:lrac-noise-data} illustrates the final distribution of noise data used for training the baseline models.

For the Track 1 baseline model, no data augmentation is applied; the model is trained exclusively on the curated speech data. Training inputs are extracted as sliding windows of 62,400 samples per utterance with 50\% overlap. In contrast, the Track 2 baseline employs on-the-fly data augmentation using the EnhPreprocessor class from the ESPnet framework. Reverberation is applied with a probability of 0.5, and additive noise with a probability of 0.8, with signal-to-noise ratios (SNRs) uniformly sampled between -5 and 30 dB. Since EnhPreprocessor supports only a single noise source per utterance, we adopt that constraint. For reverberation, we exclusively use real room impulse responses (RIRs) from the public datasets provided in the LRAC Challenge and do not include synthetically generated RIRs. When reverberation is added to an input utterance, the early reflection component of the room impulse response is also applied to the reference speech. The early reflections are defined as the 50 ms segment following the direct path.

\section{MODEL ARCHITECTURES}
\label{sec:model}

The ESPnet repository includes implementations of several neural audio codecs, including Soundstream~\cite{Soundstream} and Encodec~\cite{encodec}, from which we derive our baseline systems for the LRAC Challenge. These codecs follow a convolutional encoder-decoder architecture with a quantizer in the middle. Both Soundstream and Encodec employ a Residual Vector Quantizer (RVQ) to compress encoder embeddings. Our baseline systems adopt this design and they are trained with a generative adversarial network (GAN) approach.

\subsection{Track 1 Baseline Model}
\label{ssec:model_track1}

The Track 1 baseline model employs an encoder operating directly on the raw audio waveform. It begins with a convolutional input layer (kernel size 7, 8 output channels), followed by four convolutional blocks. Each block consists of three residual convolutional sub-blocks and a strided convolution for temporal downsampling. The block strides are 3, 4, 4, and 5, yielding an overall stride of 240 samples (10 ms). Within each residual sub-block, two dilated convolutions with ELU nonlinearities are wrapped by skip connections, enabling a larger receptive field. All convolutional layers use weight normalization. The embedding dimension increases progressively to 16, 32, 64, and finally 160 after each strided convolution. To minimize computational cost, the encoder omits a dedicated output layer. The third block includes two center-aligned convolutions, introducing 10 ms latency; combined with encoder buffering, this results in 20 ms total transmit-side latency. The encoder receptive field spans 14,085 samples.

RVQ is applied with 6 layers, each containing 1,024 codewords, contributing 10 bits per frame. With an encoder frame rate of 100 Hz, this corresponds to 1 kbps per layer, or 6 kbps in total. Each RVQ layer uses projection layers to reduce the 160-dimensional encoder output to 12 dimensions, and then project the selected codeword back to 160 dimensions, with residuals computed in the original space. The RVQ complexity is 19.35 MFLOPS. Post-training, the output projections can be absorbed into the codebooks, storing separate transmit- and receive-side versions, which reduces complexity to 17.05 MFLOPS at the cost of increased binary size.

The decoder is a convolutional network consisting of four blocks and a final output layer. Each block begins with a transposed convolution for upsampling, followed by three residual sub-blocks. The strides of the transposed convolutions are 5, 4, 3, and 4, yielding an overall stride of 240. The kernel sizes are set equal to the strides, preventing implicit overlap-add in the transposed convolutions that could otherwise introduce additional latency. As in the encoder, each residual sub-block contains two dilated convolutions with ELU nonlinearities, wrapped by skip connections. The final output layer is a convolution with kernel size 21 and a tanh activation, producing waveform samples in the range [-1, 1]. All convolutional layers use weight normalization. The decoder introduces 10 ms algorithmic latency due to the center-aligned convolution in the first block, and its overall computational complexity is 296.8 MFLOPS (excluding nonlinearities).

We provide a summary of the latency and computational complexity of the Track 1 Baseline system in Table~\ref{tab:lrac-track1-summary}. We also provide a detailed design sheet for both Track 1 and Track 2 baseline models with all the hyperparameters, latency and computational complexity calculations in \cite{lrac-design-sheet}. For a guideline on buffering and algorithmic latency calculations, see the guidance on the challenge website \cite{latency-guideline}.

\begin{table}[t]
\centering
\caption{Latency and computational complexity of the Track~1 baseline system.}
\label{tab:lrac-track1-summary}
\resizebox{0.5\textwidth}{!}{%
\begin{tabular}{lcccc}
\toprule
 & \multicolumn{2}{c}{Transmit Side} & Receive Side & Overall \\
\cmidrule(lr){2-3} \cmidrule(lr){4-4}
 & Encoder & RVQ & Decoder &  \\
\midrule
Buffering Latency (ms)     & 10    & 0    & 0    & 10 \\
Algorithmic Latency (ms)   & 10    & 0    & 10   & 20 \\
Compute Complexity (MFLOPS)& 377.5 & 17.05& 296.8& 691.35 \\
\bottomrule
\end{tabular}
}
\label{tab:latency-complexity}
\end{table}

\subsection{Track 2 Baseline Model}
\label{ssec:model_track2}

The Track 2 baseline model follows the same architectural principles as Track 1 but is trained as a joint codec and enhancement network. It takes noisy and reverberant audio as input and aims to reconstruct the clean reference signal. For reverberant inputs, the clean reference retains the early reverberation components.

The encoder starts with a convolutional input layer (kernel size 7, 8 output channels), followed by five convolutional blocks. Each block contains multiple residual convolutional sub-blocks and a strided convolution for temporal downsampling. The first two blocks include 4 residual sub-blocks each, while the last three contain 3 sub-blocks. The block strides are 2, 2, 3, 4, and 5, resulting in an overall stride of 240 samples (10 ms). Within each residual sub-block, two dilated convolutions with ELU activations are wrapped by skip connections. The embedding dimension increases progressively to 16, 32, 64, 128, and 320 after each strided convolution. The encoder exhibits a buffering latency of 10 ms, an algorithmic latency of 20 ms due to center-aligned convolutions, and a total computational complexity of 1944.2 MFLOPS.

Similar to the Track 1 system, we employ a 6-layer RVQ, with each layer containing 1,024 codewords, contributing 10 bits per frame. Each layer first projects the 320-dimensional encoder output to a 24-dimensional space, selects a codeword, and then projects it back to 320 dimensions, with residuals computed in the original space. The RVQ has a computational complexity of 48 MFLOPS. By absorbing the output projections into the codebooks after training, the complexity can be reduced to 38.7 MFLOPS at the expense of increased binary size.

The Track 2 decoder is a convolutional network composed of five blocks followed by a final output layer. Each block starts with a transposed convolution for upsampling, followed by three residual sub-blocks. The strides of the transposed convolutions are 5, 4, 3, 2, and 2, resulting in an overall stride of 240. Kernel sizes match the strides, as in the Track 1 decoder. The embedding dimension decreases progressively to 96, 48, 24, 12, and 8 after each transposed convolution. The final output layer is a convolution with a kernel size of 21, a single output channel, and a tanh activation, producing waveform samples in the range [–1, 1]. The decoder introduces 20 ms of algorithmic latency due to two center-aligned convolutions in the first block and has an overall computational complexity of 563.3 MFLOPS (excluding nonlinearities).

\begin{table}[t]
\centering
\caption{Latency and computational complexity of the Track~2 baseline system.}
\label{tab:lrac-track2-summary}
\resizebox{0.5\textwidth}{!}{%
\begin{tabular}{lcccc}
\toprule
 & \multicolumn{2}{c}{Transmit Side} & Receive Side & Overall \\
\cmidrule(lr){2-3} \cmidrule(lr){4-4}
 & Encoder & RVQ & Decoder &  \\
\midrule
Buffering Latency (ms)     & 10    & 0    & 0    & 10 \\
Algorithmic Latency (ms)   & 20    & 0    & 20   & 40 \\
Compute Complexity (MFLOPS)& 1944.2 & 38.7& 563.3& 2546.2 \\
\bottomrule
\end{tabular}
}
\label{tab:latency-complexity}
\end{table}

We provide a summary of the latency and computational complexity of the Track 2 Baseline system in Table~\ref{tab:lrac-track2-summary}. For a detailed description of the hyperparameters, as well as the latency and computational complexity calculations for both Track 1 and Track 2 baseline models, please refer to the design sheet~\cite{lrac-design-sheet}.

\begin{table*}[t]
\centering
\caption{Objective evaluation results for Track 1 baseline under clean, noisy, and reverberant conditions.}
\label{tab:lrac-objective-results}
\resizebox{\textwidth}{!}{%
\begin{tabular}{lccccc|ccccc|ccccc}
\toprule
\multirow{2}{*}{Bitrate} & \multicolumn{5}{c|}{Clean} & \multicolumn{5}{c|}{Noisy} & \multicolumn{5}{c}{Reverberant} \\
\cmidrule(lr){2-6} \cmidrule(lr){7-11} \cmidrule(lr){12-16}
 & sheet\_ssqa & scoreq\_ref & audiobox\_AE\_CE & utmos & pesq 
 & sheet\_ssqa & scoreq\_ref & audiobox\_AE\_CE & utmos & pesq
 & sheet\_ssqa & scoreq\_ref & audiobox\_AE\_CE & utmos & pesq \\
\midrule
1 kbps & 1.84 & 1.15 & 3.90 & 1.44 & 1.15 
       & 1.72 & 1.29 & 3.40 & 1.33 & 1.11
       & 1.85 & 1.36 & 2.94 & 1.26 & 1.07 \\
6 kbps & 3.84 & 0.35 & 5.28 & 3.23 & 2.67 
       & 3.12 & 0.82 & 4.37 & 2.70 & 1.81
       & 2.22 & 1.13 & 3.43 & 1.32 & 1.18 \\
\bottomrule
\end{tabular}%
}
\end{table*}

\begin{table*}[t]
\centering
\caption{Objective evaluation results for Track 2 baseline under clean, noisy, and reverberant conditions.}
\label{tab:lrac-track2-objective-results}
\resizebox{\textwidth}{!}{%
\begin{tabular}{lccccc|ccccc|ccccc}
\toprule
\multirow{2}{*}{Bitrate} & \multicolumn{5}{c|}{Clean} & \multicolumn{5}{c|}{Noisy} & \multicolumn{5}{c}{Reverberant} \\
\cmidrule(lr){2-6} \cmidrule(lr){7-11} \cmidrule(lr){12-16}
 & sheet\_ssqa & scoreq\_ref & audiobox\_AE\_CE & utmos & pesq 
 & sheet\_ssqa & scoreq\_ref & audiobox\_AE\_CE & utmos & pesq
 & sheet\_ssqa & scoreq\_ref & audiobox\_AE\_CE & utmos & pesq \\
\midrule
1 kbps & 2.07 & 1.01 & 3.96 & 1.37 & 1.21 
       & 1.95 & 1.15 & 3.70 & 1.35 & 1.18
       & 2.43 & 1.12 & 3.55 & 1.32 & 1.15 \\
6 kbps & 3.55 & 0.43 & 5.25 & 2.97 & 2.13 
       & 2.92 & 0.75 & 4.6 & 2.56 & 1.73
       & 2.67 & 0.92 & 4.25 & 1.79 & 1.29 \\
\bottomrule
\end{tabular}%
}
\end{table*}

\section{TRAINING}
\label{sec:typestyle}

We train both systems end-to-end using a combination of adversarial and reconstruction losses. The RVQ codebooks are updated with exponential moving averages, while the projection matrices are optimized via backpropagation. For the codebooks, straight-through gradient estimation is applied. We use Euclidean distance in codeword selection. To stabilize training and prevent rapid fluctuations in the encoder embeddings and codeword selections, we include a commitment loss~\cite{vq-vae}. During training, we uniformly sample between 1 kbps and 6 kbps using random quantizer dropout, enabling the decoder to operate robustly at both bitrates.

For reconstruction, we employ a multi-scale mel spectrogram loss~\cite{yamamoto} with window lengths of 64, 128, 256, 512, 1024, and 2048 samples, and corresponding mel bin counts of 10, 20, 40, 80, 160, and 320, respectively.

The adversarial objective follows Encodec~\cite{encodec}, using multi-scale feature discriminators operating in the complex STFT domain. We compute STFTs with window lengths of 128, 256, 512, 1024, and 2048 samples, with hop sizes equal to one quarter of the window length. Each discriminator is a convolutional network with weight normalization and Leaky ReLU activations (slope 0.1), using 16 channels in its internal layers. Hinge loss is applied at the output layer. In addition, we apply a feature matching loss on the intermediate discriminator representations.

The loss weights are set to 10 for the commitment loss, 5 for the multi-scale mel-spectrogram loss, 1 for the adversarial loss, and 2 for the feature matching loss.

Each training epoch consists of 10,000 randomly selected utterances from the training set. From these utterances, sliding windows of 62,400 samples are extracted with 50\% overlap. Training within an epoch continues until all windows are consumed, so the number of iterations per epoch is not fixed but remains approximately constant. We reserve 1,000 utterances for validation. For Track 2, on-the-fly noise and reverberation augmentation is applied during validation. Although offline augmentation of the validation set could help reduce variance in the validation losses, we did not adopt this approach for simplicity.

We train with a batch size of 64 per GPU, using distributed data parallelism with 6 GPUs for Track 1 and 8 GPUs for Track 2. The learning rate is initialized at 3e-4 and decays at each step by a factor of 0.998. Optimization is performed with RAdam, using betas of 0.9 and 0.999. The Track 1 model is trained for 1,150 epochs and the Track 2 model for 1,325 epochs. For model selection, we use the checkpoint with the lowest multi-scale mel-spectrogram loss on the validation set. While this choice prioritizes ease of implementation, more robust strategies—such as combining objective metrics that correlate better with subjective listening tests—are likely to yield improved results.

We present the baseline results for Track 1 and Track 2 on the open test set in Table~\ref{tab:lrac-objective-results} and Table~\ref{tab:lrac-track2-objective-results}, respectively. The reported metrics are the official objective measures of the LRAC challenge: SHEET\_SSQA, SCOREQ\_Ref, Audiobox\_AE\_CE, UTMOS, and PESQ. Further details on the open test set and these evaluation metrics are available on the 2025 LRAC Challenge objective evaluation page~\cite{objective-evaluation}.

\section{ACKNOWLEDGEMENTS}
\label{sec:typestyle}

We thank the Data Team at Cisco Collaboration AI for their support in curating and augmenting the training datasets used in the baseline development. In particular, we acknowledge the contributions of Ivana Balic, Laura Lechler, Daniel Arismendi, Ayoub Zaidour, and James Taylor.



\bibliographystyle{unsrt}
\bibliography{LRAC_baseline}

\end{document}